\documentclass[aps,prl,twocolumn]{revtex4}
\usepackage{graphicx}
\usepackage{color}
\usepackage{amsmath}
\usepackage{amssymb}
\usepackage[normalem]{ulem}
\begin{document}

\title{Protein sequence and structure: Is one more fundamental than the other?}

\author{Jayanth R. Banavar}
\affiliation{Department of Physics,
University of Maryland, College Park, MD 20742
}

\author{Trinh X. Hoang}
\affiliation{
Institute of Physics, Vietnam Academy of Science and
Technology, 10 Dao Tan, Ba Dinh, Hanoi 10000, Vietnam
}

\author{Flavio Seno, Antonio Trovato and Amos Maritan}
\affiliation{
Dipartimento di Fisica `G. Galilei', Universit\`a di Padova \&
CNISM, unit\`a di Padova, Via Marzolo 8, 35131
Padova, Italy
}

\date{Received: date / Accepted: date}

\begin{abstract}
We argue that protein native state structures reside in a novel ``phase" of
matter which confers on proteins their many amazing characteristics. This phase
arises from the common features of all globular proteins and is characterized
by a sequence-independent free energy landscape with relatively few low energy
minima with funnel-like character. The choice of a sequence that fits well into
one of these predetermined structures facilitates rapid and cooperative
folding. Our model calculations show that this novel phase facilitates the
formation of an efficient route for sequence design starting from random
peptides.
\end{abstract}

\maketitle

\section{Introduction}

There has been great progress in our understanding of inanimate matter. How is
living matter different from inanimate matter?  Our focus is on taking a fresh
look at the hints provided by Nature and develop a framework for understanding
proteins \cite{Creighton}, which are the engines of life. We suggest that (1)
protein-like structures lie in a marginally compact ``phase'' of matter in the
proximity of a transition to the swollen phase; (2) the limited menu of protein
folds \cite{Chothia1} arises independent of sequence specificity thereby
relegating the role of sequence to picking from this menu the best-fit fold for
its native state - the specific amino acids in the vicinity of the active site
also play a key role in functionality; (3) sequence design is relatively
straightforward
starting from random peptides, whose low energy conformations have snippets of
secondary motifs; (4) the native fold of a well designed sequence is robust to
mutations as is observed in experiments \cite{Creighton}; and (5) protein
structure is more fundamental than sequence even though they both play a
pivotal role in determining protein behavior.

Proteins \cite{Creighton} are short linear heteropolymers made up of amino
acids. Current theory notes that the interactions of the side-chains of the
amino acids with one another and with the water can be frustrating
\cite{Frauenfelder1979,Stein1985,Mezard1987}, as in spin glasses, resulting in
a rugged free energy landscape riddled with many local minima. A rugged
landscape is not conducive
to rapid folding, because the chain can easily get stuck in spurious local
minima and be unable to surmount the barriers required to escape from them.
Bryngelson and Wolynes \cite{Bryngelson} suggested that the choice of a
sequence, for which these frustration effects are minimized, is crucial -- the
compatibility between interacting amino acids in the native state is maximized
resulting in a funnel-like landscape \cite{Onuchic1992,Wolynes1995,Dill1997}
with few, if any, significant minima competing with the native state basin
\cite{Woly}.

The limited number of protein folds \cite{Chothia1} house a much larger number
of protein sequences in their native state.  The functionality of a protein is
largely controlled by its low energy modes of motion \cite{Bahar}.  The
interactions of proteins with other proteins and its binding partners are
heavily based on geometry. The equilibrium fluctuations around the native state
are mainly determined by the native state structure and not by the amino acid
sequence. The shape of a protein in its folded state determine the probable
motions which, in turn, directly impact on function. The chicken-egg question
that this poses is which is more fundamental: sequence or structure?  If
sequence choice is made by the principle of minimal frustration
\cite{Bryngelson}, what determines the limited menu of folds?

Strikingly, all protein native state structures are made up of the same
building blocks of helices and sheets, independent of amino acid sequence, with
hydrogen bonds providing the scaffolding \cite{Pauling1,Pauling2}. Also, the idea that
steric avoidance \cite{Ramachandran} promotes helices and sheets is a
sequence-independent result. Furthermore, distinct sequences can adopt the same
native state fold and multiple protein functionalities can arise within the
context of a single fold \cite{Creighton}. The existence of a menu of folds
largely determined by the common features of all proteins rather than sequence
specificity would make the role of protein sequence much less onerous than in
the standard picture. The chemistry of the amino acid side-chains would then be
instrumental in selecting the best-fit native fold from the pre-determined menu
of folds.

From everyday experience, the helix is a natural, compact conformation of a
short, flexible tube \cite{unitube}.
It has been shown that the helix appears as a winning
conformation of a short flexible tube under various mechanims that promote
compactness \cite{MaritanNature,KamienScience,Buried-area,Dietrich}.
The tube is anisotropic, i.e. locally the symmetry is cylindrical instead of
being spherical. One may imagine a tube to
be the continuum limit of a discrete chain of discs or coins placed
perpendicular to the chain axis. Unlike a chain of spheres, a {\em chain of
discs} accurately captures the {\em symmetry} of a chain molecule because,
associated with each point along the chain, there is a special axis, defined by
the tangent to the chain, perpendicular to the face of the disc.  Indeed, the
side chains of the amino acids stick out in a direction lying approximately
perpendicular to the tangent to the chain.  The protein-like tube
\cite{MaritanNature,HoangPNAS,HoangPNAS06,Banavar07} is distinct from an ordinary garden hose in that
the protein backbone snakes along the axis of the tube and the space within the
tube
ensures that there are no  steric overlaps.  The crucial feature is the
symmetry of the monomers or the amino acids, which can be captured either by
coin-like objects or, as shown in Ref. \cite{nanomachinePNAS}, by requiring
that the interaction between a pair of basic constituents depends not only on
their mutual distance but also on the relative orientation of their local
reference frames as determined by their neighbors along the chain.

Here, through new calculations, we reassess the role played by structures and sequences
in light of previous studies \cite{HoangPNAS,HoangPNAS06,Banavar07}.
First, we show that there are distinctions between the structures found in
the tube model and those in simple lattice models. Though the latter have
elucidated important aspects of protein folding at the cost of coarse-graining, we find 
that lattice models do not allow for the emergence
of the marginally compact phase unlike the tube model. Second, we show that
random HP sequences in the tube model have ground states with a high
content of secondary structures, just as for homopolymers and designed HP
sequences. Random mutations of a designed sequence can
occasionally switch the designed structure to a new structure, which still is in
the marginally compact phase. This suggests that the marginally compact phase
provides an efficient route for sequence design starting from random sequences
of amino acids. The role of sequence is to select structures that are
appropriate for functionality and to enhance the stability of folded structures.

\section{Models and Methods}

Two models of polymer are considered in our study. The first  \cite{Davide}
corresponds to a discrete tube homopolymer chain characterized by
just tube-like self-avoidance and a pairwise attraction between the
monomers.  The second  \cite{HoangPNAS} is more sophisticated and is aimed
at mimicking proteins. It satisfies the tube constraint and also has specific interactions
such as bending energy, pairwise hydrophobic contact interaction, and
directional hydrogen bonding with energetic and geometrical constraints.

In the first model, the polymer is modeled as a chain of $N$
beads, each of hard-core radius $R_{hc}$. The bead spacing along the chain is
equal to 1. Between any pair of non-consecutive monomers
there is an attraction in the form of a square-well potential of range
$R_{int}$. For any triplet of beads, ($i$,$j$,$k$), one can draw a circle
of radius $R_{ijk}$ going through the positions of the beads. Let
$\Delta$ be the radius of a self-avoiding tube whose axis is defined by the
positions of the beads. The tube constraint is imposed by requiring that
$R_{ijk} \geq \Delta$ for every triplet ($i$,$j$,$k$)
\cite{Gonzalez,BanavarJSP}.

In the second model, the amino acids are coarse-grained as 
beads located at the positions of the $C_\alpha$ atoms, and  placed
along the axis of a self-avoiding tube of thickness
$\Delta=2.5\AA$. The bead spacing along the chain is
3.8$\AA$. Additionally, sterics requires that two non-consecutive
$C_\alpha$'s cannot be closer than 4$\AA$ from each other. The bond
angle associated with three consecutive $C_\alpha$ atoms is
constrained to stay between $82^o$ and $148^o$. The energy
of a chain conformation is given by:
\begin{equation}
E = E_{bending} + E_{hydrophobic} + E_{hbonds} \ ,
\end{equation}
where the three terms on the right hand side correspond to bending energy,
hydrophobic energy, and hydrogen bonding energy, respectively.
The bending energy is equal to the sum of local bending penalties along the
chain. A bending penalty energy $e_R=0.3\epsilon > 0$ is applied when the local
radius of curvature at a given bead is smaller than 3.2$\AA$ (the unit
$\epsilon$ corresponds to the energy of a local hydrogen bond). The
hydrophobic energy is the total energy of all pairwise hydrophobic contacts
between amino acids. A contact is said to be formed when two non-consecutive beads are
found within a distance of 7.5$\AA$. In a homopolymer chain of amino acids,
the contact energies are all the same and equal to $e_W=-0.1\epsilon$. For
hydrophobic-polar (HP) sequences, only contacts between hydrophobic residues
are favorable and are assigned an energy of $e_{HH}=-0.5\epsilon$ per contact.
Contacts involving polar residues are given zero energy. Hydrogen bonds
have to satisfy a set of distance and angular constraints \cite{HoangPNAS} on
the $C_\alpha$s as found by a statistical analysis of native protein
structures \cite{PRE} from the PDB. A local hydrogen
bond is said to form between residues that are separated by three peptide bonds
along the chain, and is assigned an energy $-\epsilon$. A non-local hydrogen
bond is assigned an energy of $-0.7\epsilon$. Additionally, a
cooperative energy of $-0.3\epsilon$ is given for each pair of hydrogen
bonds that are formed by pairs of consecutive amino acids in the sequence.

We employ a parallel tempering \cite{Swendsen} Monte Carlo scheme for obtaining
the ground state as well as other equilibrium characteristics of the system.
For each system, 20 to 24 replicas are considered, each evolving at its own
selected temperature $T_i$. For each replica, the simulation is carried out
with standard pivot and crankshaft move sets and the Metropolis algorithm for
move acceptance. In a pivot move, one randomly chooses a bead $i$ and rotates
the shorter part of the chain (either from 1 through $i-1$ or from $i+1$ to
$N$) by a small angle and about a randomly chosen axis that goes through the
bead $i$.  In a crankshaft move, two beads $i$ and $j$ are chosen randomly such
that $|i-j|<6$, and the beads between $i$ and $j$ are rotated by a small angle
and about the axis that goes through $i$ and $j$. In both move sets, the
rotation angle is drawn randomly from a Gaussian distribution of zero mean and
a dispersion of 4$^o$. An attempt to exchange replicas is made every 100 MC
steps. The exchange of replicas $i$ and $j$ is accepted with a probability
equal to $p_{ij}=\min\{ 1, \exp[k_B^{-1}(T_i^{-1}-T_j^{-1})(E_i - E_j)] \}$,
where $k_B$ is the Boltzmann constant, and $E_i$ and $E_j$ are the energies of
the replicas at the time of the exchange.  The weighted multiple-histogram
technique \cite{Ferrenberg} is used to compute the specific heat of the system.

\section{Lattice vs. tube picture}

Simplistic lattice models \cite{Dill} have provided an useful tool to address
fundamental questions on protein folding and sequence design.
For example, it has been found in the HP model that only a few sequences
can have an unique ground state \cite{Yue1995}, and correspondingly, only
a few structures are highly designable \cite{HaoLi}.
Consider a lattice model of a chain made up of two
kinds of monomers, H and P, representing hydrophobic and polar
propensities. Typically one assumes an effective attraction
between non-bonded neighboring H-H pairs and no other
interactions otherwise.  For a homopolymer comprising a chain
of H monomers, all compact conformations are degenerate ground
states. The degeneracy grows with chain length.  One is able to
optimally design the sequence of a heteropolymer in order to
ensure that it has a unique ground state which is likely to be
but is not necessarily maximally compact -- the choice of the
sequence removes the large degeneracy.

\begin{figure}
\center
\includegraphics[width=3in]{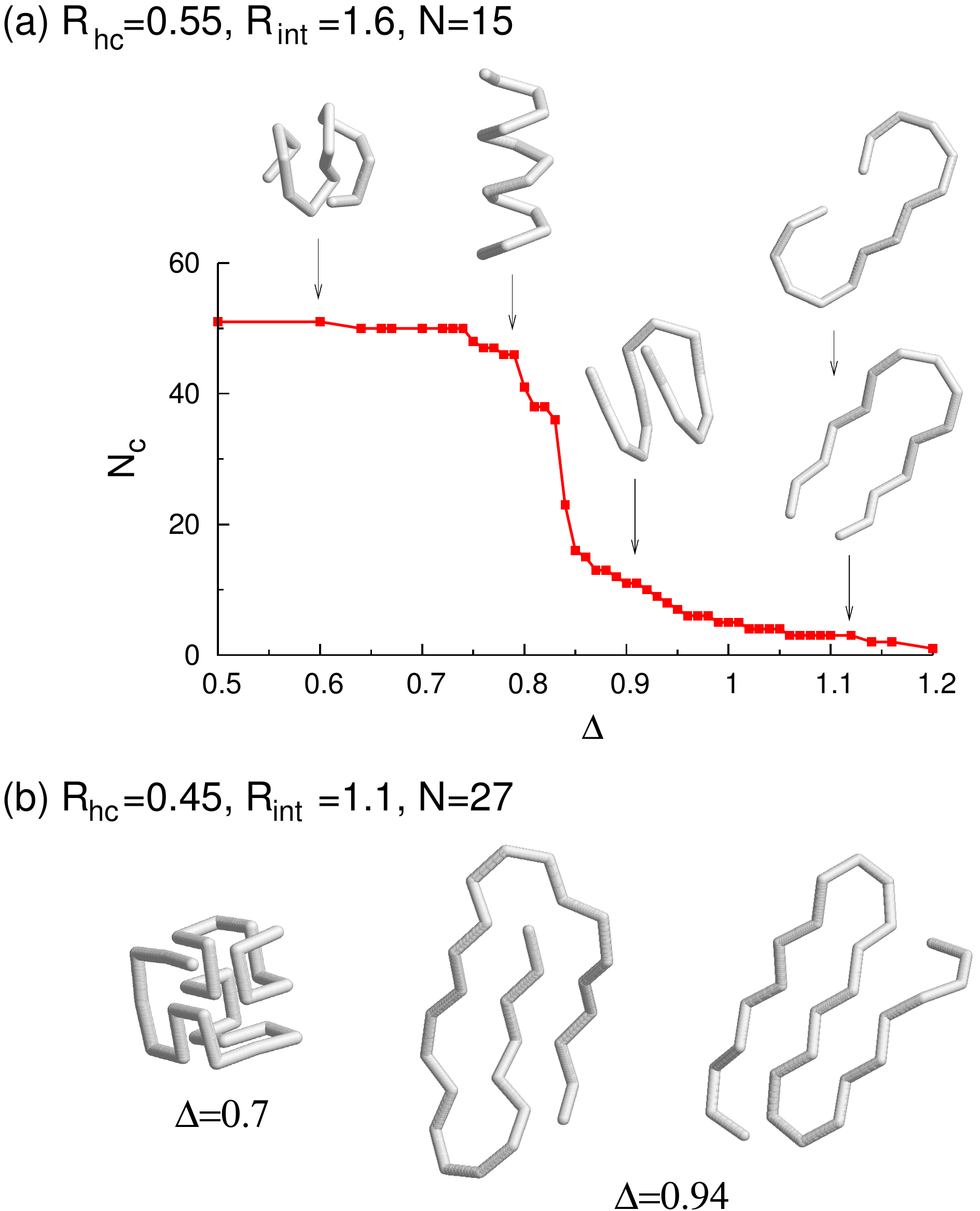}
\caption{
Discrete homopolymer in the tube picture.
(a) Dependence of maximal number of contacts, $N_c$, on the tube thickness,
$\Delta$, for a short homopolymer of $N=15$ beads. $R_{hc}$ is the hard-core
radius of the beads and $R_{int}$ is the range that defines a contact
between two non-consecutive beads. The figure shows a decreasing function
of $N_c$ in multiple steps as $\Delta$ increases. Ground state conformations
are shown for several values of $\Delta$ corresponding to the end of some
plateaus as indicated.
(b) Ground state conformations of a chain of $N=27$ beads
for selected values of $R_{hc}$, $R_{int}$ and $\Delta$ as indicated. The
conformations shown include a lattice-like 3x3x3 maximally compact conformation
and two types of planar sheets.
}
\end{figure}

In contrast, for a tube, on varying the thickness, one goes from a compact
phase to a marginally compact phase with relatively small degeneracy
arising from the constraint that spatially nearby tube segments
must lie parallel to each other as in helices and
sheets.
Fig. 1 shows results of calculations for a discrete homopolymer in the tube
picture \cite{Davide}. It has been shown that in the absence of the tube
constraint, one gets compact conformations with significant degeneracy on
maximizing the number of contacts \cite{Davide}.
For the set of parameters shown in Fig. 1a,
we have determined the maximum number of possible contacts that the polymer can
have as a function of the tube thickness. The number of contacts is found to
decrease in discrete steps as the tube thickness increases. We have plotted the
ground state conformation at the end of each plateau corresponding to the
conformation with the largest thickness (and therefore the greatest wiggle
room) that has number of contacts equal to the plateau value. There is a sharp
drop in the number of contacts as maximally compact conformations, for small
values of the thickness,  give way to marginally compact conformations as the
thickness increases and ultimately yields the swollen phase (with few or no
contacts) for large values of the thickness. These marginally compact
conformations include the helix and the hairpin. The key point is that there is
a thinning of the number of degenerate conformations in the marginally compact
phase and the marginally compact conformations include the building block
motifs of protein structures. This is the case even for a homopolymer.
Interestingly, for more finely tuned parameters and chain length one can get
compact conformations of a simple cubic lattice (Fig. 1b). Here, in the compact
phase, one obtains all maximally compact 3x3x3 conformations. Again on
increasing the thickness, one enters the marginally compact phase - two of the
conformations in this phase are also shown in the figure and are small sheets
made up of zig-zag strands.
Our calculations show that though lattice conformations can be
obtained in the tube picture, they do not belong to the marginally compact
phase of a flexible tube characterized by a remarkable low degeneracy of
ground state conformations.

Let's consider now a homopolymer in a tube model of proteins \cite{HoangPNAS}.
It has been
shown \cite{HoangPNAS} that, by changing $e_R$ and $e_W$, one finds a marginally
compact phase of ground states and low-lying energy minimum conformations that
are protein-like. It is suggested that structures in this phase constitute
a menu of pre-determined folds that a protein sequence can
choose from.  Fig. 2 shows the temperature dependence of
the specific heat for a homopolymer of length $N=48$ amino acids with
parameters $e_R$ and $e_W$ chosen such that the ground state is
a three-helix bundle.
The specific heat shows multiple transitions from
the swollen phase at high temperatures to a collapsed phase of marginally
compact structures at low temperatures. One finds that the competing energy
minima to the ground state are all characterized by protein-like tertiary
structures. Interestingly, the $\beta$-sheets are found to be present more
frequently at higher temperatures.

The above result is obtained for a homopolymer and suggests that geometry
and symmetry are responsible for the selection of putative native state
structures even before the sequence has had a chance to weigh in. Thus sequence
design becomes easier and the sequence has a less onerous task in sculpting a
folding funnel landscape compared to the HP lattice model, where the sequence
must not only break the degeneracy of the maximally compact structures but also
create a folding funnel. There are distinct advantages for protein native state
structures to be at the edge of compactness. In the vicinity of a phase
transition (note that we are dealing with modest sized systems and the
transition will necessarily be rounded), the system would be expected to be
exquisitely sensitive to the right kinds of perturbations conferring the
amazing functionalities that proteins possess.

\begin{figure}
\center
\includegraphics[width=3.2in]{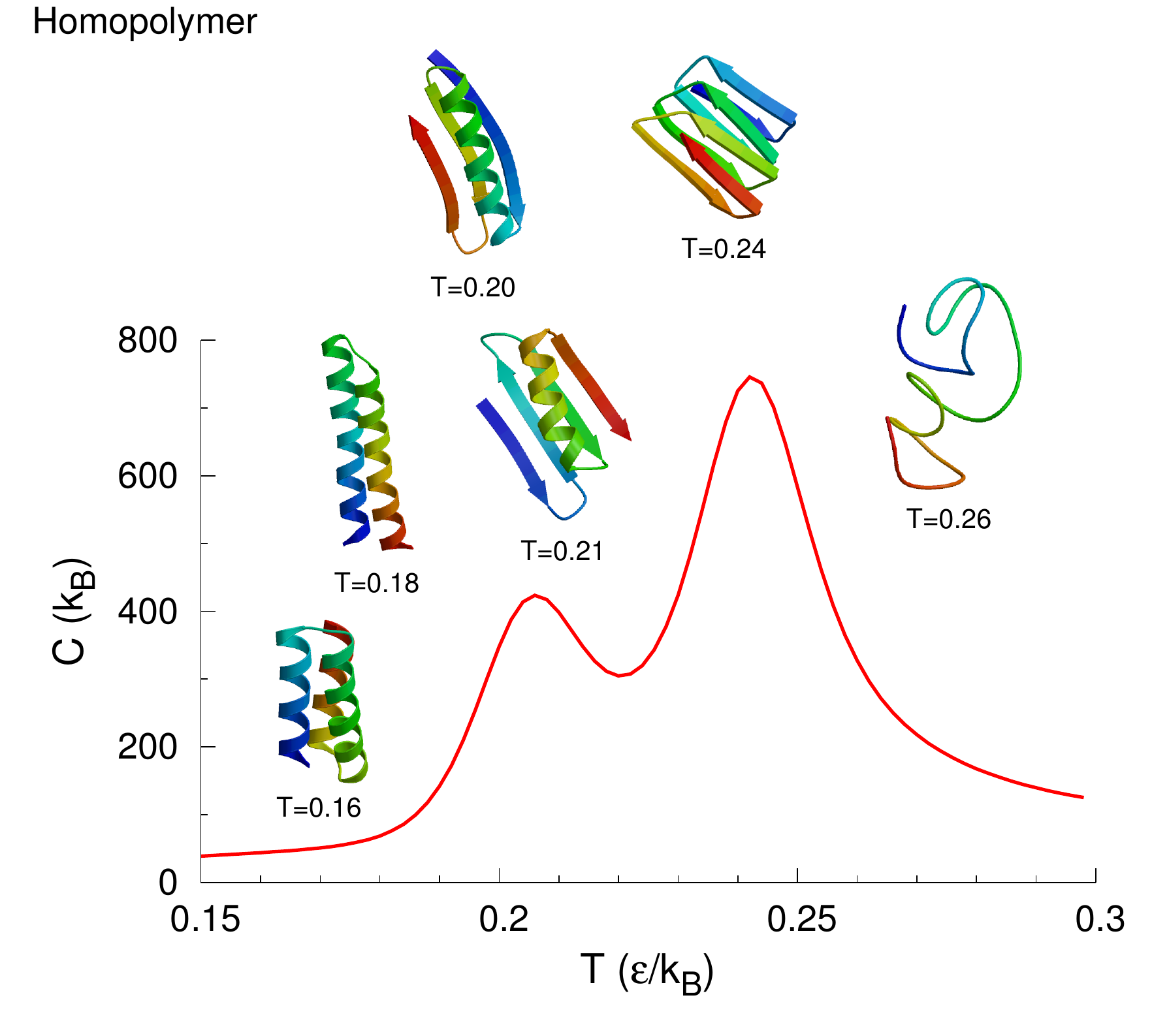}
\caption{
Temperature dependence of the specific heat, $C$, of a homopolymer of length of
$N=48$ beads in the tube model of proteins. The model parameters are the
bending energy penalty, $e_R=0.3\varepsilon$, and a favorable hydrophobic
energy $e_W=-0.1\varepsilon$ between nearby beads within a distance of 7.5$\AA$
from each other, where $\varepsilon$ denotes the magnitude of the energy of a
local hydrogen bond.  The data are obtained through parallel tempering Monte
Carlo simulations and by using the weighted histogram technique.  The peak of
the specific heat at the higher temperature corresponds to a transition from
the swollen phase to the marginally compact phase.  The peak at the lower
temperature corresponds to a transition between competing marginally compact
conformations. Helical conformations are favored at lower temperatures, whereas
a structure comprised of $\beta$-sheets is preferred at higher temperatures.
The ground state of the system is the three-helix bundle.  Representative
conformations at various temperatures are shown.
}
\end{figure}

\section{Designed vs. random sequences}

Consider now the tube model with just two types of amino acids, hydrophobic (H)
and polar (P), in which pairwise attraction is given only between the H
residues.  In the marginally compact phase, it is relatively easy to design a
sequence that folds to a specific structure.
It was shown by Hoang et al. \cite{HoangPNAS06} that there are relatively simple
recipes in the design procedure to get a fragment of the chain to form a
helix or a $\beta$-sheet. Specifically, a fragment of periodic patterns like
HPPH or HPPPH (the H residues are separated by 2 or 3 P residues) is likely to
form a helix. In contrast, a fragment of pattern like HPHPH (the H residues are
separated by one P residue) is likely to from a $\beta$-sheet. Interestingly,
these recipes are consistent with the successful experimental design of {\it de novo} proteins and amyloid-like fibrils \cite{Hecht}.

Fig. 3b shows a three-helix bundle folded by a designed HP sequence. Folding of
this sequence is highly cooperative as demonstrated by a sharp peak in the
specific heat (Fig. 3c).  In order to further assess the role of sequences, we
ask how good is the folding of random sequences compared to that of the
designed sequence? We found that, in the marginally compact phase, random
sequences also have ground states characterized by a high content of secondary
structures (Fig. 3a).  However, the designed sequence has significantly higher
stability and folding cooperativity manifested by the position and the height
of the specific heat peak respectively (Fig. 3c).  The designed sequence folds
with much greater ease than the random heteropolymers.  At low temperatures,
the specific heat of the designed sequence is smaller than for random
sequences, highlighting that, in the former case, there is a unique ground
state well separated in energy from other excited states.

Next, we proceed to assess the robustness of designed sequences against random
mutations of amino acids. For this purpose, we designed a 24-bead HP sequence
(PPPHHPPHHPPPPHPHPPPPHPHP) that folds to a zinc-finger motif (Fig.  4a).
Mutations are made from H to P or vice versa. For each mutated sequence, the
ground state and the equilibrium properties are calculated through parallel
tempering Monte Carlo simulations. We found that in 17 out of 24 single point
mutations, the ground state conformation does not change. These mutations
include all positions in the helix region (residues 1-12) and several positions
in the $\beta$-hairpin. Other single point mutations (at positions
14,17,18,19,20,21,23) in the $\beta$-hairpin region completely destabilize this
structure and convert it into a helix or a loop.  Mutations also are likely to
change the height and position of the specific heat, and thus affect the
folding properties of the sequence either to poorer or better (Fig. 4b).  We
also tried a limited number of double and triple random mutations of the
original sequence and found that the zinc-finger structure can persist in about
50\% of double mutations but in none of the triple mutations considered.

The above results suggest that there is an efficient route for an
evolutionary sequence design \cite{Shakh} starting from a random sequence. In
the marginally compact phase of a chain molecule, structures are stable enough
so that a single point mutation on a very short 24 amino acid chain usually
does not
destroy the folded state. Yet a few-point mutation may switch the chain to a
new conformation which can be more
functionally useful. Without mutational stability, it would be very hard
for an evolutionary selected sequence to survive.

\begin{figure}
\center
\includegraphics[width=3in]{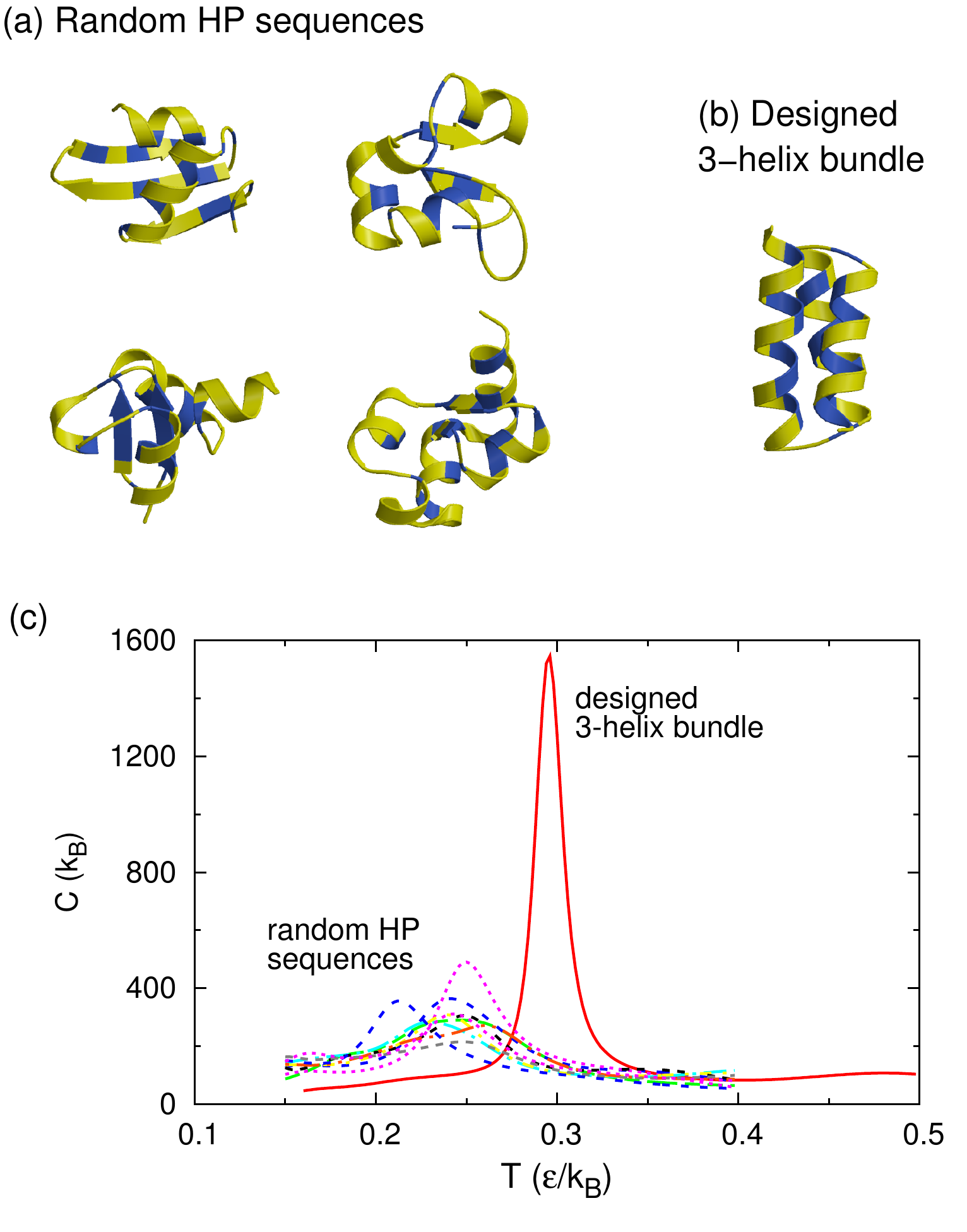}
\caption{
Random and designed HP sequences in the tube model of
proteins. (a) Ground state conformations of several random HP sequences of
$N=48$ beads. The hydrophobic (H) and polar (P) amino acids are shown in blue
and yellow respectively. The hydrophobic interaction is present only between
nearby  H amino acids and is assigned an energy $e_{HH} = -0.5\varepsilon$ in
the model. The bending energy penalty is uniform for all amino acids and is
equal to $e_R=0.3\varepsilon$. (b) Ground state of a designed HP sequence that
folds into a three-helix bundle. The parameters chosen are the same as for the
sequences in (a). (c) Temperature dependence of the specific heat, $C$, for 10
random HP sequences (dashed and dotted lines) and for the designed three-helix
bundle (solid line). The data are obtained through parallel tempering Monte
Carlo simulations. }
\end{figure}

\begin{figure}
\center
\includegraphics[width=3.2in]{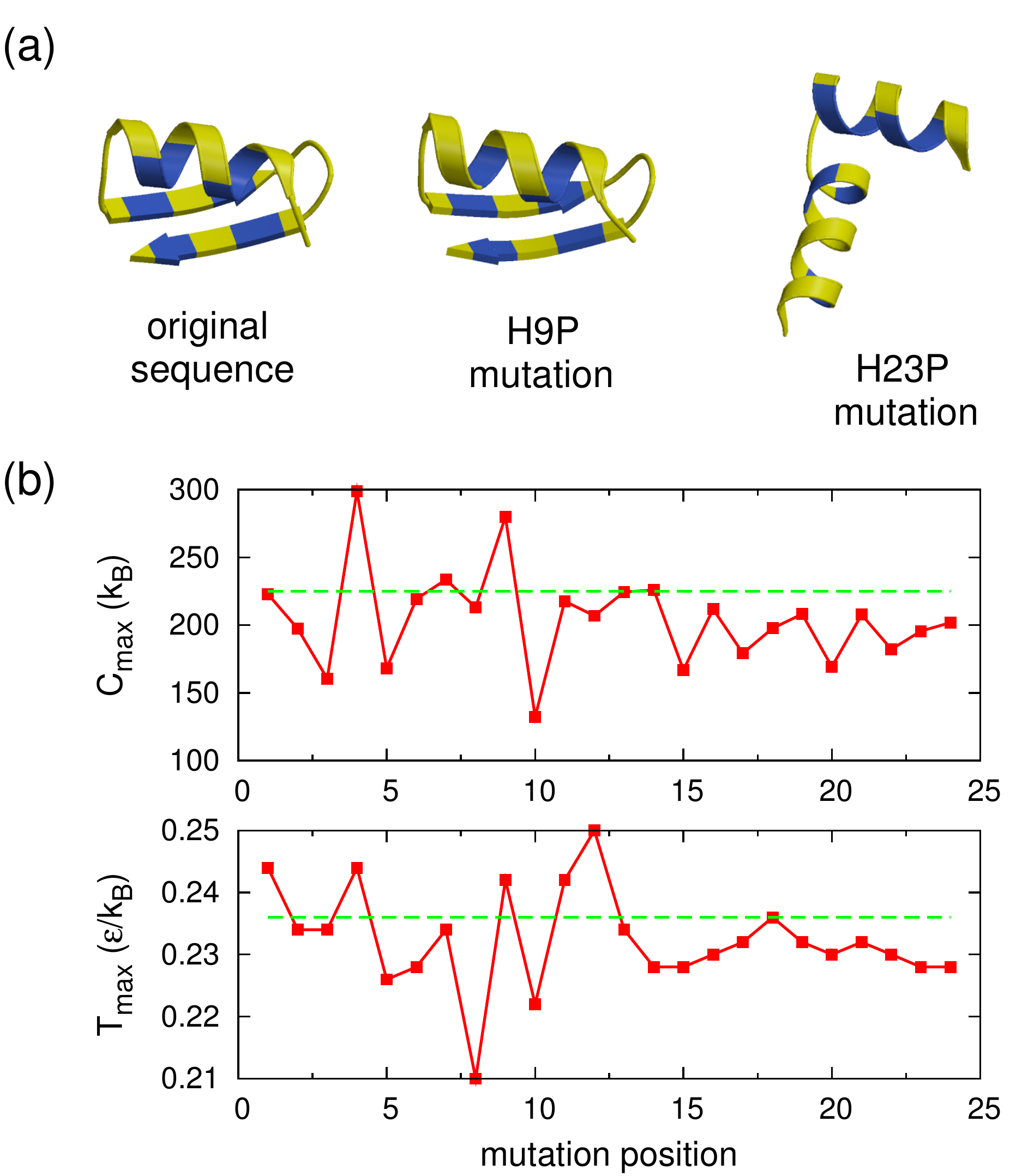}
\caption{
Mutations of a designed HP sequence. (a) Ground state conformations of the
original sequence and two mutated sequences at positions 9 and 23 as indicated.
In 17 out of 24 single point mutations the ground state does not change. (b)
Dependence of the maximum of the specific heat, $C_{max}$, and its temperature
of occurrence, $T_{max}$, on mutation position for 24 possible single mutations
of the original sequence (solid lines). Values of $C_{max}$ and $T_{max}$ for
the original sequence are indicated by horizontal dashed lines. Higher
$T_{max}$ indicates a higher stability whereas higher $C_{max}$ indicates a
higher folding cooperativity.
}
\end{figure}

\section{Discussion}

The tube picture not only provides an
elegant explanation for the novel phase selected by Nature to house
biomolecular structures but also bridges this phase and conventional polymer
phases on reducing the tube thickness. Additionally, upon increasing the length
of the chain molecule or the number of chains, one observes, in computer
simulations, a crossover to semi-crystalline structures with different portions
of the backbone chain lying parallel to one another as extensively
verified in \cite{Auer-prl2008,Auer-prl2010}.
Significantly, this low temperature anisotropic phase of tubes provides a
simple rationalization of the formation of amyloid in mis-folded proteins
\cite{KellyPrusinerDobson} (leading to deadly diseases, including Alzheimer's
and the Mad Cow disease) and the formation of semicrystalline polymer phases
\cite{semicrystal,Bassett1981,Strobel1997}.

The number of ground state structures in the marginally compact phase of a tube
is much smaller than the corresponding number for chains of spheres: the energy
landscape is vastly simpler. Second, the resulting structures are marginally
compact (the effects of attractive self-interactions have just set in) and,
because of their proximity to a phase transition to the swollen phase, are
sensitive to the right type of perturbations.  Because protein sequences are
necessarily required to select from a limited menu of folds \cite{Chothia1},
there is a many-to-one mapping from sequences to structures. Indeed, as
observed experimentally, the native state fold is often robust to mutations in
which one amino acid is changed into another in accord with experimental
observations \cite{BakerZZYY}. Sequences evolve rapidly without any deleterious
consequences in terms of functionality, because the native state fold remains
the same and continues to be able to have structure-based interactions with
other proteins and cell-products. Also, two proteins evolutionarily related to
each other are likely to share the same fold. The successful interpretation of
dynamical experiments \cite{Creighton} and their sensitivity to amino acid
mutations follows naturally from these observations. Interestingly, not all
pre-sculpted structures are necessarily chosen by natural proteins as their
native states \cite{Laio}.

A great simplicity in understanding inanimate matter is the concept of phases.
The key point is that the gross properties of a material are often determined
by the phase of matter that the material resides in.  One might wonder whether
living matter has adopted a powerful strategy by poising the native state
structures of proteins in a novel marginally compact phase of matter. This
would suggest that this phase could be exploited in the laboratory for the
creation of powerful nanomachines \cite{nanomachinePNAS} and artificial life by
networking these machines to yield novel emergent behavior.

We are grateful to Ken Dill, Bob Jernigan, George Rose, and Peter
Wolynes for helpful discussions. We acknowledge support from
Fondazione Cariparo, from Vietnam National Foundation for Science and
Technology Development (NAFOSTED) grant 103.01-2010.11, and from
Programmi di Ricerca Scientifica di Rilevante Interessa Nazionale
Grant SKNEWA in 2009.


\begin{thebibliography}{99}
\bibitem{unitube}
A hard sphere can be thought of as a point particle or a zero
dimensional object with its own private space of spatial extent
equal to its radius.  Generalizing to a one dimensional object,
one must consider a line or a string, with private space
associated with each point along the line, leading to a uniform
tube with its axis defined by the line.
\end{thebibliography}
\end{document}